\documentclass[allclo]{FBSart}
\usepackage{amsfonts}
\usepackage{amssymb}
\usepackage{graphicx} 

\newcommand{\sfrac}[2]{\mbox{\footnotesize $\displaystyle \frac{#1}{#2}$}}
 
\newcommand{\lsim}{\mathrel{\rlap{\lower4pt\hbox{\hskip0pt$\sim$}} 
\raise1pt\hbox{$<$}}}           
\newcommand{\gsim}{\mathrel{\rlap{\lower4pt\hbox{\hskip0pt$\sim$}} 
\raise1pt\hbox{$>$}}}           

\title{Sigma Terms of Light-Quark Hadrons}

\author{V.\,V. Flambaum,\instnr{1,3} A.\ H\"oll,\instnr{2} 
P. Jaikumar,\instnr{2} C.\,D.\ Roberts\instnr{2,4} and S.\,V.\ Wright\instnr{2}} 

\instlist{
Argonne Fellow, Physics Division, Argonne National Laboratory, Argonne, IL
60439-4843, USA\and
Physics Division, Argonne National Laboratory, Argonne, IL
60439-4843, USA
\and
School of Physics, University of New South Wales, Sydney 2052, Australia
\and
Institut f\"ur Physik, Universit\"at Rostock, D-18051 Rostock, 
Germany}

\runningauthor{V.\,V. Flambaum, et al.}
\runningtitle{Sigma Terms of Light-Quark Hadrons}
\begin{document}

\maketitle 
\begin{abstract}
A calculation of the current-quark mass dependence of hadron masses can help in using observational data to place constraints on the variation of nature's fundamental parameters.  A hadron's $\sigma$-term is a measure of this dependence.
The connection between a hadron's $\sigma$-term and the Feynman-Hellmann theorem is illustrated with an explicit calculation for the pion using a rainbow-ladder truncation of the Dyson-Schwinger equations: in the vicinity of the chiral limit $\sigma_\pi = m_\pi/2$.  
This truncation also provides a decent estimate of $\sigma_\rho$ because the two dominant self-energy corrections to the $\rho$-meson's mass largely cancel in their contribution to $\sigma_\rho$.  
The truncation is less accurate for the $\omega$, however, because there is little to compete with an $\omega \to \rho \pi$ self-energy contribution that magnifies the value of $\sigma_\omega$ by $\lesssim 25$\%.
A Poincar\'e covariant Faddeev equation, which describes baryons as composites of confined-quarks and -nonpointlike-diquarks, is  solved to obtain the current-quark mass dependence of the masses of the nucleon and $\Delta$, and thereby $\sigma_N$ and $\sigma_\Delta$.  This ``quark-core'' piece is augmented by the ``pion cloud'' contribution, which is positive.  The analysis yields $\sigma_N \simeq 60\,{\rm MeV}$ and $\sigma_\Delta \simeq 50\,{\rm MeV}$.
%




\end{abstract}

\section{Introduction}
Chiral symmetry is explicitly broken in QCD by the current-quark mass term, which for the $u$- and $d$-quark sector is expressed in the action as 
\begin{eqnarray}
\int d^4z\, \bar Q(z) \, {\cal M}\,  Q(z) 
& = & \int d^4z\, (\bar u(z) \, \bar d(z))
\left(\begin{array}{cc}
m_u & 0 \\
0 & m_d \end{array}\right) 
\left( \begin{array}{c}
u(z)\\
d(z)
\end{array}\right) \\
& = & \int d^4z\, \left\{ \bar m\, \bar Q(z) \tau^0 Q(z) + \bar Q(z) \check m \tau^3 Q(z)\right\}, \label{eq2}
\end{eqnarray}
where: $(\tau^0)_{ij}=\delta_{ij}$ and $\{\tau^k;k=1,2,3\}$ are Pauli matrices; and $\bar m = (m_u + m_d)/2$ and $\check m = (m_u - m_d)/2$.  Empirical success with the application of chiral effective theories to low-energy phenomena in QCD indicates that this term can often be treated as a perturbation.  That simplification owes fundamentally to the phenomenon of dynamical chiral symmetry breaking (DCSB) in QCD; namely, the feature that the dressed-quark Schwinger function is nonperturbatively modified at infrared momenta: $p\lesssim 1\,$GeV.  This is a longstanding prediction of Dyson-Schwinger equation (DSE) studies \cite{lane,politzer,cdragw,bastirev,alkoferrev} that has recently been verified in numerical simulations of lattice-regularised QCD \cite{latticequarkbowman}.  A quantitative comparison and feedback between DSE and lattice studies is currently proving fruitful; e.g., Refs.\ \cite{fischer,bhagwat,maris,mandarvertex,bhagwat2,jisvertex,kurt,MP,alex}.

The $\sigma$-term for a hadron $H$ is obtained from the isoscalar matrix element 
\begin{equation}
\label{sigmaX}
\langle H(x)| \bar m\,J_\sigma(z)|H(y) \rangle\,, \; J_\sigma(z)=\bar Q(z) \tau^0 Q(z)\,,
\end{equation}
and simple counting of field dimensions entails that for mesons the scalar form factor associated with this matrix element has mass-dimension two, while for fermions it has mass-dimension one.  For all hadrons the $\sigma$-term vanishes in the chiral limit.  It is thus a keen probe of the impact of explicit chiral symmetry breaking on a hadron, in particular, as will be made plain below, on a hadron's mass.

Such information is important for numerous reasons, some of longstanding \cite{gls90}, but our interest is prompted by its connection with the variation of nature's fundamental parameters \cite{flambaum,nollett,uzan}.  It is a feature anticipated of models for the unification of all interactions that the so-called fundamental ``constants'' actually exhibit spatial and temporal variation.  In consequence there is an expanding search for this variation via laboratory, astronomical and geochemical measurements.  An interpretation of some of these measurements can benefit from calculations of the current-quark mass dependence of the parameters characterising nuclear systems.  

Of relevance herein are meson and baryon masses.  A variation in light-meson masses will modify the internucleon potential, and a variation in the nucleon mass will affect the kinetic energy term in the nuclear Hamiltonian.  Such changes could modify the binding energy in deuterium.  That would have a material impact on Big Bang Nucleosynthesis (BBN) because the first step in BBN is the process $p+n \to d +\gamma$.  The rate of this process is crucially dependent on the binding energy of deuterium, and this reaction is the seed for all subsequent processes and therefore the primordial abundance of light elements.  A calculation of the current-quark mass dependence of hadron properties is therefore necessary to enable the use of observational data to place constraints on the variation of nature's constants \cite{Dmitriev}.  
 
A variation of meson and nucleon masses will also modify the position of compound resonances in heavy nuclei \cite{shuryak}.  The position two-billion years ago of such a resonance in neutron capture by $^{149}$Sm has been determined using data from the Oklo natural nuclear reactor, with a non-zero shift reported cf.\ the present day  \cite{Oklo}.  The position of this resonance is very sensitive to the masses of mesons and the nucleon, and thus the Oklo data provide an acute method by which to measure the temporal rate of variation of nature's basic parameters.  Data on the position of nuclear resonances several billion years ago are also available from the study of nuclear reactions in stars \cite{shuryak}.  Moreover, the interpretation of numerous measurements of quasar absorption spectra and superprecise atomic clocks in terms of the variation of nature's fundamental parameters also requires calculations of the quark-mass-dependence of meson and nucleon properties \cite{shuryak,clocks,tsa}.  These observations emphasise the important role that the calculations reported herein can currently play in the interpretation of many measurements performed in several areas of physics and astronomy.  In addition, a large number of new and more accurate measurements are soon expected to appear.

In Sect.\,\ref{LMesons} we focus on the $\sigma$-term for mesons, in particular, the $\pi$, $\rho$ and $\omega$.  We present an explicit calculation of $\sigma_\pi$ from the pion's scalar form factor in the rainbow-ladder truncation of the DSEs.  This is a useful means by which to exemplify the Feynman-Hellmann theorem in the context of QCD.  In order to compute the $\sigma$-terms of the nucleon and $\Delta$ it is first necessary to have a tool with which to calculate the masses of these baryons.  Hence, in Sect.\,\ref{LBaryons} we recapitulate on a Poincar\'e covariant Faddeev equation that serves this purpose \cite{arneJ}.  In Sect.\,\ref{LResults} we describe how this equation is used to calculate the dressed-quark contribution to $\sigma_N$ and $\sigma_\Delta$, and how that result may be affected by meson-cloud corrections.  Section \ref{LEpilogue} is an epilogue.  It contains a tabulation of our ``best estimates'' for $\delta m_H/m_H$ in terms of $\delta \bar m/\bar m$.

\section{Mesons}
\label{LMesons}
We judge that it is useful to begin with an explicit calculation of $\sigma_\pi$ from the pion's scalar form factor, which is defined by a momentum space re-expression of Eq.\,(\ref{sigmaX}):
\begin{equation}
\label{sigmapiQ}
s_\pi(Q^2)=\langle \pi(P^\prime)| \bar m\, J_\sigma(Q)|\pi(P) \rangle\,, \; Q_\mu=(P^\prime - P)_\mu\,.
\end{equation}
This three-point function arises, for example, in the analysis of $\pi \pi$ scattering, as explicated in Ref.\,\cite{mariscotanch}.

In the rainbow-ladder truncation of QCD's DSEs,
\begin{eqnarray}
\nonumber
s_\pi(Q^2) &=& {\rm tr}_{CDF} \int\! \frac{d^4\ell}{(2\pi)^4} \,{\cal S}(\ell_{-1,\frac{1}{2}})\,\bar m \, \Gamma_{\tau^0}(\ell_{-1,0};Q)\,{\cal S}(\ell_{-1,-\frac{1}{2}})\, \\
&& \times \Gamma_\pi(\ell_{-\frac{1}{2},0};P^\prime)\, {\cal S}(\ell_{0,\frac{1}{2}})\, \Gamma_\pi(\ell_{-\frac{1}{2},\frac{1}{2}};P)\,,
\label{sigmaRL}
\end{eqnarray}
where the trace is over colour, flavour and spinor indices, and \mbox{$\ell_{\alpha,\beta}= \ell + \alpha P + \beta Q$}.  The rainbow-ladder approximation is the first term in a nonperturbative, systematic and symmetry preserving DSE truncation scheme \cite{munczek,truncscheme}.  This leading-order contribution preserves the one-loop renormalisation group properties of QCD, and has provided a uniformly accurate description and prediction of a wide range of meson properties \cite{marisrev}.  (NB.\ We employ the Euclidean metric described in \protect\ref{App:EM}~Euclidean Conventions.)

Equation (\ref{sigmaRL}) is fully renormalised.  It contains the renormalised rainbow-dressed quark propagator, which is the solution of 
\begin{eqnarray}
{\cal S}(p)^{-1} & =&  Z_2 \,(i\gamma\cdot p + {\cal M}^{\rm bm}) + \Sigma(p)\,, \label{gendse} \\
\Sigma(p) & = & Z_1 \int^\Lambda_q\! g^2 D_{\mu\nu}(p-q) \frac{\lambda^a}{2}\gamma_\mu {\cal S}(q) \frac{\lambda^a}{2}\gamma_\nu , \label{gensigma}
\end{eqnarray}
wherein: $\int^\Lambda_q$ represents a Poincar\'e invariant regularisation of the integral, with $\Lambda$ the regularisation mass-scale; $D_{\mu\nu}$ describes the interaction between light-quarks; and ${\cal M}^{\rm bm}$ describes the $\Lambda$-dependent current-quark bare masses.  The final step in any calculation is to take the limit $\Lambda\to\infty$.  The quark-gluon-vertex and quark wave function renormalisation constants, $Z_{1,2}(\zeta,\Lambda)$, depend on the gauge parameter, the renormalisation point, $\zeta$, and the regularisation mass-scale, but they are independent of the current-quark mass.  Equation~(\ref{sigmaRL}) is supplemented by the renormalisation condition
\begin{equation}
\label{renormS} \left.{\cal S}(p)^{-1}\right|_{p^2=\zeta^2} = i\gamma\cdot p +
{\cal M}(\zeta)\,,
\end{equation}
where ${\cal M}(\zeta)$ describes the renormalised (running) current-quark masses: 
\begin{equation}
Z_2(\zeta,\Lambda) \, {\cal M}^{\rm bm}(\Lambda) = Z_4(\zeta,\Lambda) \, {\cal M}(\zeta)\,,
\end{equation}
with $Z_4(\zeta,\Lambda)$ the Lagrangian-mass renormalisation constant.  

Equation (\ref{sigmaRL}) contains in addition: the Bethe-Salpeter amplitude for the ladder-pion bound state, which is obtained from
\begin{equation}
\Gamma_\pi(k;P) = - \int^\Lambda_q\! g^2 D_{\mu\nu}(k-q) \frac{\lambda^a}{2}\gamma_\mu {\cal S}(q_+))\Gamma_\pi(q;P) {\cal S}(q_-) \frac{\lambda^a}{2}\gamma_\nu , 
\end{equation}
with $q_\pm = q \pm P/2$, and normalised canonically 
\begin{eqnarray}
\nonumber
2 P_\mu &= &  {\rm tr}_{CDF}\int^\Lambda_q\! \Gamma_\pi(q;-P) \frac{\partial}{\partial P_\mu} \,{\cal S}(q+P/2) \Gamma_\pi(q;P) {\cal S}(q-P/2) \\
&& + \,{\rm tr}_{CDF}\int^\Lambda_q\! \Gamma_\pi(q;-P)  {\cal S}(q+P/2) \Gamma_\pi(q;P) \frac{\partial}{\partial P_\mu}\,{\cal S}(q-P/2);
\label{canonicalnorm}
\end{eqnarray}
and the isoscalar vertex
\begin{equation}
\Gamma_{\tau^0}(\ell;Q) =  Z_4(\zeta,\Lambda) \tau^0 
 + \int^\Lambda_q\! g^2 D_{\mu\nu}(\ell-q) \frac{\lambda^a}{2}\gamma_\mu {\cal S}(q_+) \,  \Gamma_{\tau^0}(\ell;Q)  {\cal S}(q_-) \frac{\lambda^a}{2}\gamma_\nu \,.
\end{equation}
NB.\ Although the isoscalar vertex $\Gamma_{\tau^0}(\ell;Q)$ depends on the renormalisation point, the product $\bar m(\zeta)\,\Gamma_{\tau^0}(\ell;Q)$ is renormalisation-point-independent.

The pion $\sigma$-term is defined via
\begin{eqnarray}
2 \, m_\pi \, \sigma_\pi & :=& s_\pi(Q^2=0)\\
\nonumber
&=&{\rm tr}_{CDF} \int\! \frac{d^4\ell}{(2\pi)^4} \,{\cal S}(\ell_{-1,0})\,\bar m \, \Gamma_{\tau^0}(\ell_{-1,0};0)\,{\cal S}(\ell_{-1,0})\, \\
&& \times \Gamma_\pi(\ell_{-\frac{1}{2},0};-P)\, {\cal S}(\ell)\, \Gamma_\pi(\ell_{-\frac{1}{2},0};P)\,. \label{sigmaA}
\end{eqnarray}
Equation (\ref{sigmaA}) can be simplified using a textbook result; viz.,
\begin{equation}
\frac{\partial}{\partial \bar m(\zeta)} \, {\cal S}(k) = -\, {\cal S}(k)\, \Gamma_{\tau^0}(k;0) \, {\cal S}(k)\,,
\end{equation}
and hence
\begin{eqnarray}
\nonumber \lefteqn{2 \, m_\pi \,\sigma_\pi 
= - \,\bar m(\zeta)\,{\rm tr}_{CDF} \int\! \frac{d^4\ell}{(2\pi)^4} \,\frac{\partial {\cal S}(\ell_{-1,0})}{\partial \bar m(\zeta)}\, \,\Gamma_\pi(\ell_{-\frac{1}{2},0};-P)\, {\cal S}(\ell)\, \Gamma_\pi(\ell_{-\frac{1}{2},0};P)} \\
\nonumber& = & - \,\bar m(\zeta) \frac{\partial P_\mu}{\partial \bar m(\zeta)} {\rm tr}_{CDF} \int\! \frac{d^4\ell}{(2\pi)^4}  \frac{\partial {\cal S}(\ell_{-1,0})}{\partial P_\mu}\,\Gamma_\pi(\ell_{-\frac{1}{2},0};-P)\, {\cal S}(\ell)\, \Gamma_\pi(\ell_{-\frac{1}{2},0};P)\\
\nonumber
& = &  - \, \bar m(\zeta) \frac{\partial P^2}{\partial \bar m(\zeta)} = \bar m(\zeta) \frac{\partial m_\pi^2}{\partial \bar m(\zeta)}\,,
\end{eqnarray}
where the last line follows from Eq.\,(\ref{canonicalnorm}), so that   
\begin{equation}
\sigma_\pi = \bar m(\zeta) \frac{\partial m_\pi}{\partial \bar m(\zeta)} \,. \label{sigmapiend}
\end{equation}

We emphasise that in arriving at Eq.\,(\ref{sigmapiend}) we have depended heavily upon the fact that the rainbow-ladder expression is the leading term in a systematic, nonperturbative and symmetry preserving truncation of the DSEs \cite{mandarvertex,munczek,truncscheme}.  The derivation provides a concrete illustration of a general result that may be viewed as a consequence of the Feynman-Hellmann theorem.  Applied in the present case, this theorem states that the response of an eigenvalue of the QCD mass-squared operator to a change in a parameter in that operator is given by the expectation value of the derivative of the mass-squared operator with respect to the parameter.  The result is valid in this form for all mesons; i.e.,
\begin{equation}
\label{sigmameson}
2\, m_M \sigma_{M} := s_M(0) = \bar m(\zeta) \frac{\partial m_M^2}{\partial \bar m(\zeta)} \; \Rightarrow\; \sigma_{M} = \bar m(\zeta) \frac{\partial m_M}{\partial \bar m(\zeta)} \,.
\end{equation}
NB.\ The $\sigma$-term is a renormalisation point invariant, in general and also in our explicit calculation.

In QCD the pion's mass is expressed precisely via \cite{mrt98}:
\begin{equation}
\label{gmor}
m_\pi^2 = -\,2\, \bar m(\zeta) \frac{\rho_\pi(\zeta)}{f_\pi}\,.
\end{equation}
In this expression, $f_\pi$ is the pion's leptonic decay constant: 
\begin{equation} 
\label{fpi0} f_{\pi} \,\delta^{ij} \,  P_\mu = Z_2(\zeta,\Lambda)\,{\rm tr} \int^\Lambda_q 
\sfrac{1}{2} \tau^i \gamma_5\gamma_\mu\, {\cal S}(q_+) \Gamma^j_{\pi}(q;P) 
{\cal S}(q_-)\,;
\end{equation} 
and $\rho_{\pi}$ is the residue of the pion pole in the dressed-quark-antiquark-pseudoscalar vertex:\footnote{NB.\ The factor of $Z_2(\zeta,\Lambda)$ in Eq.\,(\ref{fpi0}) guarantees that the right-hand-side (r.h.s.) is gauge invariant, independent of the renormalisation point and finite in the limit $\Lambda \to \infty$; and $Z_4(\zeta,\Lambda)$ in Eq.\,(\ref{cpres}) ensures that the r.h.s.\ is gauge invariant, finite when the regularisation scale is removed, and evolves according to the renormalisation group precisely as required to guarantee that the product $\bar m(\zeta) \rho_{\pi}\!(\zeta)$ is renormalisation point independent.} 
\begin{equation} 
\label{cpres} i  \rho_{\pi}\!(\zeta)\, \delta^{ij}  = Z_4(\zeta,\Lambda)\,{\rm tr} 
\int^\Lambda_q \sfrac{1}{2} \tau^i \gamma_5 \, {\cal S}(q_+) 
\Gamma^j_{\pi}(q;P) {\cal S}(q_-)\,.
\end{equation} 

Additional information may now be obtained from Eq.\,(\ref{sigmapiend}) by substituting Eq.\,(\ref{gmor}).  In the neighbourhood of the chiral limit \cite{mrt98}
\begin{equation}
\rho_{\pi}\!(\zeta) \stackrel{m\sim 0}{=} - \frac{\langle \bar q q \rangle^0_\zeta}{f_\pi^0},
\end{equation}
where $\langle \bar q q \rangle^0_\zeta$ is the vacuum quark condensate and $f_\pi^0$ is the chiral limit value of the leptonic decay constant, and hence
\begin{equation}
\label{sigmamassratiopion}
2 \, m_\pi \, \sigma_\pi \stackrel{\bar m \sim 0}{=} - 2 \,\bar m(\zeta)\, \frac{\langle \bar q q \rangle^0_\zeta}{(f_\pi^0)^2} \; \Rightarrow \; \sigma_\pi \stackrel{\bar m \sim 0}{=} \frac{1}{2}\, m_\pi\,.
\end{equation}
This is an essential consequence of DCSB.

With typical, calculated chiral limit values of \cite{marisrev}: $f_\pi^0 = 0.088\,$GeV and $\langle \bar q q \rangle^0_{\zeta=1\,{\rm GeV}} = (-\,0.241\,{\rm GeV})^3$, and a current-quark mass $m(\zeta=1\,{\rm GeV})= 0.0055\,$GeV, one estimates from Eq.\,(\ref{sigmamassratiopion})
\begin{equation}
\sigma_\pi = 0.95\,\surd m(\zeta=1\,{\rm GeV}) = 0.071\,{\rm GeV}\,.
\end{equation}
On the other hand, the value obtained via direct calculation in rainbow-ladder truncation can be determined from the information provided in Refs.\,\cite{mr97,maristandyrho,marisvienna} using Eq.\,(\ref{sigmapiend}); namely,
\begin{equation}
\sigma_\pi^{RL}=0.069\,{\rm GeV}.
\end{equation}


It is useful and relevant to compare the pion $\sigma$-term with that obtained for the $\rho$-meson, and one may infer the rainbow-ladder truncation result from Refs.\,\cite{maristandyrho,marisvienna}; viz., 
\begin{equation}
2 \,m_\rho\,\sigma_\rho^{RL} = s_\rho^{RL}(0) = (0.194\,{\rm GeV})^2 
\; \Rightarrow \; \sigma_\rho^{RL} 
= 0.025\,{\rm GeV}\,.
\end{equation}

Since the $\rho$-meson is unstable it is natural to ask whether meson-loop self-energy corrections alter this result significantly.  It was shown in Ref.\,\cite{hawespichowsky} that the $\rho \to \pi\pi$ and $\rho\to\omega \pi$ self-energy insertions, $\Pi_\rho^{\pi\pi}$ and $\Pi_\rho^{\omega\pi}$ respectively, provide the largest shift in the mass of the $\rho$-meson.  Hence, we focus on the contribution of these terms to $\sigma_\rho$ and estimate their effect by adapting the approach of Ref.\,\cite{wrightpion}, which gives formulae for $\Pi_\rho^{\pi\pi}$ and $\Pi_\rho^{\omega\pi}$ that can be used to calculate their $m_\pi^2$-dependence.  For all values of $m_\pi$, both $\Pi_\rho^{\pi\pi}<0$ and $\Pi_\rho^{\omega\pi}<0$; viz., they both act to reduce $m_\rho^2$.  However, since the $\rho\to\pi\pi$ decay channel is open in the vicinity of the experimental value of $m_\pi$, we find
\begin{equation} 
\label{rhoself}
0 < 
\left. m_\pi^2 \frac{\partial \Pi_\rho^{\omega\pi}}{\partial m_\pi^2} 
\right|_{(m_\pi^2)_{\rm expt.}} \!\!\!\!\!\! = (0.056\,{\rm GeV})^2 \lesssim
- \left. m_\pi^2 \frac{\partial \Pi_\rho^{\pi\pi}}{\partial m_\pi^2}\right|_{(m_\pi^2)_{\rm expt.}} \!\!\!\!\!\! = (0.090\,{\rm GeV})^2 \,.
\end{equation}
Hence the contributions to $\sigma_\rho$ from these self-energy insertions cancel each other to a large degree, and one obtains
\begin{equation}
2 \,m_\rho\, \sigma_\rho^{RL+\pi\pi+\omega\pi} = (0.181\,{\rm GeV})^2 \; \Rightarrow \; \sigma_\rho^{RL+\pi\pi+\omega\pi} = 0.022\,{\rm GeV}\,.
\end{equation}
This suggests that in connection with $\sigma_\rho$ the effect of meson-loop contributions is moderate; i.e., $\lesssim 15\,$\%, so that the rainbow-ladder truncation yields a fairly good estimate.  It is thus noteworthy that with the interaction employed in Ref.\,\cite{maristandyrho} the rainbow-ladder truncation yields, via the formulae presented in Ref.\,\cite{marisvienna},
\begin{equation}
\label{sigmarhoRL}
\sigma_\rho^{RL} = 0.034\, m_\rho^{RL} \,.
\end{equation}

At this point we have sufficient information to estimate the $\omega$-meson's $\sigma$-term.  In the case of isospin symmetry, the $\omega$- and $\rho$-mesons are indistinguishable in rainbow-ladder truncation \cite{pras1} and hence
\begin{equation}
\sigma_\omega^{RL} = \sigma_\rho^{RL}\,.
\end{equation}
However, the states differ in the nature of their self-energy corrections.  In contrast with the $\rho$-meson, the $\omega \to \pi\pi$ coupling vanishes.  This leaves only the $\omega\to \rho \pi$ self-energy, and in the isosymmetric limit\footnote{NB.\ The $\omega \to \pi\pi\pi$ contribution is negligible \cite{hollenberg} and no other strong decay channels are open.  The signs are incorrect in Eqs.\,(2.11) and (3.4) of Ref.\,\cite{hollenberg}.}
\begin{equation}
\Pi_\omega^{\rho \pi} = 3\, \Pi_\rho^{\omega \pi}\,.
\end{equation}
From this and Eq.\,(\ref{rhoself}) 
\begin{equation}
\label{sigmaw}
2 \, m_\omega\, \sigma_\omega^{RL+\rho\pi} = (0.217\,{\rm GeV})^2 \; \Rightarrow \; \sigma_\omega^{RL+\rho\pi} = 0.032\,{\rm GeV} = 0.043\,m_\omega\,,
\end{equation}
which is an increase of $\lesssim 25$\% over the rainbow-ladder result.  The marked difference between the values in Eqs.\,(\ref{sigmarhoRL}) and (\ref{sigmaw}), and that in Eq.\,(\ref{sigmamassratiopion}) highlights again the particular character of QCD's Goldstone mode.

\section{Baryons}
\label{LBaryons}
In a consideration of Eq.\,(\ref{sigmaX}) and the $\sigma$-term for fermions, the analogue of Eq.\,(\ref{sigmameson}) is
\begin{equation}
\label{sigmaFermion}
\sigma_{F} := s_F(0) = \bar m(\zeta) \frac{\partial M_F}{\partial \bar m(\zeta)} \,.
\end{equation}
This is apparent in the context of the Feynman-Hellmann theorem because fermion fields have mass-dimension $3/2$ and their spectrum is described by a mass operator, not a mass-squared operator.  The $\sigma$-terms of the nucleon and $\Delta$ can thus be determined in any framework that provides for a calculation of the current-quark-mass dependence of $M_N$ and $M_\Delta$.  We will use the Poincar\'e covariant Faddeev equation model described in Refs.\,\cite{arneJ}.

\subsection{Faddeev Equation}
In order to explain the input it is useful to recapitulate on the model's foundation.  For quarks in the fundamental representation of colour-$SU(3)$:
\begin{equation} 
3_c \otimes 3_c \otimes 3_c = (\bar 3_c \oplus 6_c) \otimes 3_c = 1_c \oplus 
8_c^\prime \oplus 8_c \oplus 10_c\,,
\end{equation} 
and hence any two quarks in a colour-singlet three-quark bound state must constitute a relative colour-antitriplet.  This fact enables the derivation of a Faddeev equation for the bound state contribution to the three quark scattering kernel \cite{regfe} because the same kernel that describes mesons so well \cite{marisrev} is also attractive for quark-quark scattering in the colour-$\bar 3$ channel.  

In this truncation of the three-body problem the interactions between two selected quarks are added to yield a quark-quark scattering matrix, which is then approximated as a sum over all possible diquark pseudoparticle terms  \cite{cjbfe}: Dirac-scalar $+$ -axial-vector $+ [\ldots]$.  The Faddeev equation thus obtained describes the baryon as a composite of a dressed-quark and nonpointlike diquark with an iterated exchange of roles between the bystander and diquark-participant quarks.  The baryon is consequently represented by a Faddeev amplitude: 
\begin{equation} 
\Psi = \Psi_1 + \Psi_2 + \Psi_3 \,, 
\end{equation} 
where the subscript identifies the bystander quark and, e.g., $\Psi_{1,2}$ are obtained from $\Psi_3$ by a correlated, cyclic permutation of all the quark 
labels.  

\subsection{{\it Ans\"atze} for the Nucleon and $\Delta$}
\label{ANDelta}
We employ the simplest realistic representation of the Faddeev amplitudes for the nucleon and $\Delta$.  The spin- and isospin-$1/2$ nucleon is a sum of scalar and axial-vector diquark correlations:
\begin{equation} 
\label{Psi} \Psi_3(p_i,\alpha_i,\tau_i) = {\cal N}_3^{0^+} + {\cal N}_3^{1^+}, 
\end{equation} 
with $(p_i,\alpha_i,\tau_i)$ the momentum, spin and isospin labels of the 
quarks constituting the bound state, and $P=p_1+p_2+p_3$ the system's total momentum.\footnote{NB.\ Hereafter we assume isospin symmetry of the strong interaction; i.e., the $u$- and $d$-quarks are indistinguishable but for their electric charge.  This simplifies the form of the Faddeev amplitudes.}  Since it is not possible to combine an isospin-$0$ diquark with an isospin-$1/2$ quark to obtain isospin-$3/2$, the spin- and isospin-$3/2$ $\Delta$ contains only an axial-vector diquark component
\begin{equation}
\label{PsiD} \Psi^\Delta_3(p_i,\alpha_i,\tau_i) = {\cal D}_3^{1^+}.
\end{equation} 

The scalar diquark piece in Eq.\,(\ref{Psi}) is 
\begin{eqnarray} 
{\cal N}_3^{0^+}(p_i,\alpha_i,\tau_i)&=& [\Gamma^{0^+}(\sfrac{1}{2}p_{[12]};K)]_{\alpha_1 
\alpha_2}^{\tau_1 \tau_2}\, \Delta^{0^+}(K) \,[{\cal S}(\ell;P) u(P)]_{\alpha_3}^{\tau_3}\,,%
\label{calS} 
\end{eqnarray} 
where: the spinor satisfies (recall \protect\ref{App:EM}~Euclidean Conventions)
\begin{equation}
(i\gamma\cdot P + M)\, u(P) =0= \bar u(P)\, (i\gamma\cdot P + M)\,,
\end{equation}
with $M$ the mass obtained by solving the Faddeev equation, and it is also a
spinor in isospin space with $\varphi_+= {\rm col}(1,0)$ for the proton and
$\varphi_-= {\rm col}(0,1)$ for the neutron; $K= p_1+p_2=: p_{\{12\}}$,
$p_{[12]}= p_1 - p_2$, $\ell := (-p_{\{12\}} + 2 p_3)/3$; $\Delta^{0^+}$
is a pseudoparticle propagator for the scalar diquark formed from quarks $1$
and $2$, and $\Gamma^{0^+}\!$ is a Bethe-Salpeter-like amplitude describing
their relative momentum correlation; and ${\cal S}$, a $4\times 4$ Dirac
matrix, describes the relative quark-diquark momentum correlation.  (${\cal
S}$, $\Gamma^{0^+}$ and $\Delta^{0^+}$ are discussed in Sect.\,\ref{completing}.)  The colour antisymmetry of $\Psi_3$ is implicit in $\Gamma^{J^P}\!\!$, with the 
Levi-Civita tensor, $\epsilon_{c_1 c_2 c_3}$, expressed via the antisymmetric 
Gell-Mann matrices; viz., defining 
\begin{equation} 
\{H^1=i\lambda^7,H^2=-i\lambda^5,H^3=i\lambda^2\}\,, 
\end{equation} 
then $\epsilon_{c_1 c_2 c_3}= (H^{c_3})_{c_1 c_2}$.  [See 
Eqs.\,(\ref{Gamma0p}), (\ref{Gamma1p}).]

The axial-vector component in Eq.\,(\ref{Psi}) is
\begin{eqnarray} 
{\cal N}^{1^+}(p_i,\alpha_i,\tau_i) & =&  [{\tt t}^i\,\Gamma_\mu^{1^+}(\sfrac{1}{2}p_{[12]};K)]_{\alpha_1 
\alpha_2}^{\tau_1 \tau_2}\,\Delta_{\mu\nu}^{1^+}(K)\, 
[{\cal A}^{i}_\nu(\ell;P) u(P)]_{\alpha_3}^{\tau_3}\,,
\label{calA} 
\end{eqnarray} 
where the symmetric isospin-triplet matrices are 
\begin{equation} 
{\tt t}^+ = \frac{1}{\surd 2}(\tau^0+\tau^3) \,,\; 
{\tt t}^0 = \tau^1\,,\; 
{\tt t}^- = \frac{1}{\surd 2}(\tau^0-\tau^3)\,, 
\end{equation} 
and the other elements in Eq.\,(\ref{calA}) are straightforward generalisations of those in Eq.\,(\ref{calS}). 

The general form of the Faddeev amplitude for the spin- and isospin-$3/2$ $\Delta$ is complicated.  However, isospin symmetry means one can focus on the $\Delta^{++}$ with it's simple flavour structure, because all the charge states are degenerate, and consider 
\begin{equation}
{\cal D}_3^{1^+}= [{\tt t}^+ \Gamma^{1^+}_\mu(\sfrac{1}{2}p_{[12]};K)]_{\alpha_1 \alpha_2}^{\tau_1 \tau_2}
\, \Delta_{\mu\nu}^{1^+}(K) \, [{\cal D}_{\nu\rho}(\ell;P)u_\rho(P)\, \varphi_+]_{\alpha_3}^{\tau_3}\,, \label{DeltaAmpA} 
\end{equation} 
where $u_\rho(P)$ is a Rarita-Schwinger spinor, Eq.\,(\ref{rarita}).

The general forms of the matrices ${\cal S}(\ell;P)$, ${\cal A}^i_\nu(\ell;P)$ and ${\cal D}_{\nu\rho}(\ell;P)$, which describe the momentum space correlation between the quark and diquark in the nucleon and the $\Delta$, respectively, are described in Ref.\,\cite{oettelfe}.  The requirement that ${\cal S}(\ell;P)$ represent a positive energy nucleon; namely, that it be an eigenfunction of $\Lambda_+(P)$, Eq.\,(\ref{Lplus}), entails
\begin{equation}
\label{Sexp} 
{\cal S}(\ell;P) = s_1(\ell;P)\,I_{\rm D} + \left(i\gamma\cdot \hat\ell - \hat\ell \cdot \hat P\, I_{\rm D}\right)\,s_2(\ell;P)\,, 
\end{equation} 
where $(I_{\rm D})_{rs}= \delta_{rs}$, $\hat \ell^2=1$, $\hat P^2= - 1$.  In the nucleon rest frame, $s_{1,2}$ describe, respectively, the upper, lower component of the bound-state nucleon's spinor.  Placing the same constraint on the axial-vector component, one has
\begin{equation}
\label{Aexp}
 {\cal A}^i_\nu(\ell;P) = \sum_{n=1}^6 \, p_n^i(\ell;P)\,\gamma_5\,A^n_{\nu}(\ell;P)\,,\; i=+,0,-\,,
\end{equation}
where ($ \hat \ell^\perp_\nu = \hat \ell_\nu + \hat \ell\cdot\hat P\, \hat P_\nu$, $ \gamma^\perp_\nu = \gamma_\nu + \gamma\cdot\hat P\, \hat P_\nu$)
\begin{equation}
\begin{array}{lll}
A^1_\nu= \gamma\cdot \hat \ell^\perp\, \hat P_\nu \,,\; &
A^2_\nu= -i \hat P_\nu \,,\; &
A^3_\nu= \gamma\cdot\hat \ell^\perp\,\hat \ell^\perp\,,\\
A^4_\nu= i \,\hat \ell_\mu^\perp\,,\; &
A^5_\nu= \gamma^\perp_\nu - A^3_\nu \,,\; &
A^6_\nu= i \gamma^\perp_\nu \gamma\cdot\hat \ell^\perp - A^4_\nu\,.
\end{array}
\end{equation}
Finally, requiring also that ${\cal D}_{\nu\rho}(\ell;P)$ be an eigenfunction of $\Lambda_+(P)$, one obtains
\begin{equation}
{\cal D}_{\nu\rho}(\ell;P) = {\cal S}^\Delta(\ell;P) \, \delta_{\nu\rho} + \gamma_5{\cal A}_\nu^\Delta(\ell;P) \,\ell^\perp_\rho \,,
\end{equation}
with ${\cal S}^\Delta$ and ${\cal A}^\Delta_\nu$ given by obvious analogues of Eqs.\,(\ref{Sexp}) and (\ref{Aexp}), respectively.

One can now write the Faddeev equation satisfied by $\Psi_3$ as
\begin{equation} 
 \left[ \begin{array}{r} 
{\cal S}(k;P)\, u(P)\\ 
{\cal A}^i_\mu(k;P)\, u(P) 
\end{array}\right]\\ 
 = -\,4\,\int\frac{d^4\ell}{(2\pi)^4}\,{\cal M}(k,\ell;P) 
\left[ 
\begin{array}{r} 
{\cal S}(\ell;P)\, u(P)\\ 
{\cal A}^j_\nu(\ell;P)\, u(P) 
\end{array}\right] .
\label{FEone} 
\end{equation} 
The kernel in Eq.~(\ref{FEone}) is 
\begin{equation} 
\label{calM} {\cal M}(k,\ell;P) = \left[\begin{array}{cc} 
{\cal M}_{00} & ({\cal M}_{01})^j_\nu \\ 
({\cal M}_{10})^i_\mu & ({\cal M}_{11})^{ij}_{\mu\nu}\rule{0mm}{3ex} 
\end{array} 
\right] 
\end{equation} 
with 
\begin{equation} 
 {\cal M}_{00} = \Gamma^{0^+}\!(k_q-\ell_{qq}/2;\ell_{qq})\, 
S^{\rm T}(\ell_{qq}-k_q) \,\bar\Gamma^{0^+}\!(\ell_q-k_{qq}/2;-k_{qq})\, 
S(\ell_q)\,\Delta^{0^+}(\ell_{qq}) \,, 
\end{equation} 
where: $\ell_q=\ell+P/3$, $k_q=k+P/3$, $\ell_{qq}=-\ell+ 2P/3$, 
$k_{qq}=-k+2P/3$ and the superscript ``T'' denotes matrix transpose; and
\begin{eqnarray}
\nonumber
\lefteqn{({\cal M}_{01})^j_\nu= {\tt t}^j \,
\Gamma_\mu^{1^+}\!(k_q-\ell_{qq}/2;\ell_{qq})} \\
&& \times 
S^{\rm T}(\ell_{qq}-k_q)\,\bar\Gamma^{0^+}\!(\ell_q-k_{qq}/2;-k_{qq})\, 
S(\ell_q)\,\Delta^{1^+}_{\mu\nu}(\ell_{qq}) \,, \label{calM01} \\ 
\nonumber \lefteqn{({\cal M}_{10})^i_\mu = 
\Gamma^{0^+}\!(k_q-\ell_{qq}/2;\ell_{qq})\, 
}\\ 
&&\times S^{\rm T}(\ell_{qq}-k_q)\,{\tt t}^i\, \bar\Gamma_\mu^{1^+}\!(\ell_q-k_{qq}/2;-k_{qq})\, 
S(\ell_q)\,\Delta^{0^+}(\ell_{qq}) \,,\\ 
\nonumber \lefteqn{({\cal M}_{11})^{ij}_{\mu\nu} = {\tt t}^j\, 
\Gamma_\rho^{1^+}\!(k_q-\ell_{qq}/2;\ell_{qq})}\\ 
&&\times \, S^{\rm T}(\ell_{qq}-k_q)\,{\tt t}^i\, \bar\Gamma^{1^+}_\mu\!(\ell_q-k_{qq}/2;-k_{qq})\, 
S(\ell_q)\,\Delta^{1^+}_{\rho\nu}(\ell_{qq}) \,. \label{calM11} 
\end{eqnarray} 

The $\Delta$'s Faddeev equation is
\begin{eqnarray} 
{\cal D}_{\lambda\rho}(k;P)\,u_\rho(P) & = & 4\int\frac{d^4\ell}{(2\pi)^4}\,{\cal 
M}^\Delta_{\lambda\mu}(k,\ell;P) \,{\cal D}_{\mu\sigma}(\ell;P)\,u_\sigma(P)\,, \label{FEDelta} 
\end{eqnarray} 
with
\begin{equation}
{\cal M}^\Delta_{\lambda\mu} = {\tt t}^+ 
\Gamma_\sigma^{1^+}\!(k_q-\ell_{qq}/2;\ell_{qq})\,
 S^{\rm T}\!(\ell_{qq}-k_q)\, {\tt t}^+\bar\Gamma^{1^+}_\lambda\!(\ell_q-k_{qq}/2;-k_{qq})\, 
S(\ell_q)\,\Delta^{1^+}_{\sigma\mu}\!(\ell_{qq}). 
\end{equation}

\subsection{Completing the Faddeev Equation Kernels}
\label{completing}
To complete the Faddeev equations, Eqs.\,(\ref{FEone}) \& (\ref{FEDelta}), one must specify the dressed-quark propagator, the diquark Bethe-Salpeter amplitudes and the diquark propagators that appear in the kernels.
\subsubsection{Dressed-quark propagator} 
\label{subsubsec:S} 
The dressed-quark propagator has the general form 
\begin{eqnarray} 
S(p) & = & -i \gamma\cdot p\, \sigma_V(p^2) + \sigma_S(p^2) = 1/[i\gamma\cdot p\, A(p^2) + B(p^2)]\label{SpAB} 
\end{eqnarray}
and can be obtained from QCD's gap equation, the rainbow-ladder truncation of which is given in Eq.\,(\ref{gendse}).  It is a longstanding prediction of DSE studies in QCD that the wave function renormalisation and dressed-quark mass: 
\begin{equation} 
\label{ZMdef}
Z(p^2)=1/A(p^2)\,,\;M(p^2)=B(p^2)/A(p^2)\,, 
\end{equation} 
respectively, receive strong momentum-dependent corrections at infrared momenta \cite{lane,politzer,cdragw}: $Z(p^2)$ is suppressed and $M(p^2)$ enhanced.  The enhancement of $M(p^2)$ is central to the appearance of a constituent-quark mass-scale and an existential prerequisite for Goldstone modes.  The mass function evolves with increasing $p^2$ to reproduce the asymptotic behaviour familiar from perturbative analyses, and that behaviour is unambiguously evident for $p^2 \gtrsim 10\,$GeV$^2$ \cite{mishasvy,wrightgap}. 

The impact of this infrared dressing on hadron phenomena has long been emphasised \cite{cdrpion} and, while numerical solutions of the quark DSE are now readily obtained, the utility of an algebraic form
for $S(p)$ when calculations require the evaluation of numerous
multidimensional integrals is self-evident.  An efficacious parametrisation 
of $S(p)$, which exhibits the features described above, has been used 
extensively in hadron studies \cite{bastirev,alkoferrev,marisrev}.  It is expressed via
\begin{eqnarray} 
\bar\sigma_S(x) & =&  2\,\bar m \,{\cal F}(2 (x+\bar m^2)) + {\cal
F}(b_1 x) \,{\cal F}(b_3 x) \,  
\left[b_0 + b_2 {\cal F}(\epsilon x)\right]\,,\label{ssm} \\ 
\label{svm} \bar\sigma_V(x) & = & \frac{1}{x+\bar m^2}\, \left[ 1 - {\cal F}(2 (x+\bar m^2))\right]\,, 
\end{eqnarray}
with $x=p^2/\lambda^2$, $\bar m$ = $m/\lambda$, 
\begin{equation}
\label{defcalF}
{\cal F}(x)= \frac{1-\mbox{\rm e}^{-x}}{x}  \,, 
\end{equation}
$\bar\sigma_S(x) = \lambda\,\sigma_S(p^2)$ and $\bar\sigma_V(x) =
\lambda^2\,\sigma_V(p^2)$.  The mass-scale, $\lambda=0.566\,$GeV, and
parameter values\footnote{$\epsilon=10^{-4}$ in Eq.\ (\ref{ssm}) acts only to
decouple the large- and intermediate-$p^2$ domains.}
\begin{equation} 
\label{tableA} 
\begin{array}{ccccc} 
   \bar m& b_0 & b_1 & b_2 & b_3 \\\hline 
   0.00897 & 0.131 & 2.90 & 0.603 & 0.185 
\end{array}\;, 
\end{equation} 
were fixed in a least-squares fit to light-meson observables \cite{mark,valencedistn}.  The dimensionless $u=d$ current-quark mass in Eq.~(\ref{tableA}) corresponds to
\begin{equation} 
\label{mcq}
m=5.08\,{\rm MeV}\,. 
\end{equation} 

The parametrisation yields a Euclidean constituent-quark mass
\begin{equation} 
\label{MEq} M_{u,d}^E = 0.33\,{\rm GeV}, 
\end{equation} 
defined as the solution of $p^2=M^2(p^2)$ \cite{mr97}.  Since the dressed-quark mass function is a renormalisation point invariant then so is $M^E$.  Hence, one may define a constituent-quark $\sigma$-term
\begin{equation}
\label{sMEQ}
\sigma_Q := m(\zeta) \, \frac{\partial  M^E}{\partial m(\zeta)} = 6.2\,{\rm MeV}
\end{equation}
for the parametrisation we have just described.  NB.\ With the model interaction of Ref.\,\protect\cite{maristandyrho}, Eq.\,(\protect\ref{gendse}) gives $\sigma_Q = 9.4\,$MeV in rainbow-ladder truncation; and that of Ref.\,\cite{mandarvertex} gives $\sigma_Q \sim 8\,$-$\,9\,$MeV, depending on the \textit{Ansatz} for the quark-gluon vertex, with the lower value corresponding to rainbow-ladder truncation.  It is thus likely that the parametrisation of $S(p)$ employed herein underestimates $\sigma_Q$, owing to the implicit assumption that $b_{0,1,2,3}$ are $m$-independent.

\subsubsection{Diquark Bethe-Salpeter amplitudes}
\label{qqBSA}
The rainbow-ladder DSE truncation yields asymptotic diquark states in the strong interaction spectrum.  Such states are not observed and their appearance is an artefact of the truncation.  Higher order terms in the quark-quark scattering kernel, whose analogue in the quark-antiquark channel do not much affect the properties of vector and flavour non-singlet pseudoscalar mesons, ensure that QCD's quark-quark scattering matrix does not exhibit singularities which correspond to asymptotic diquark states~\cite{mandarvertex}.  Nevertheless, studies with kernels that do not produce diquark bound states, do support a physical interpretation of the masses, $m_{(qq)_{J^P}}$, obtained using the rainbow-ladder truncation: the quantity $l_{(qq)_{J^P}}=1/m_{(qq)_{J^P}}$ may be interpreted as a range over which the diquark correlation can persist inside a baryon.  These observations motivate the {\it Ansatz} for the quark-quark scattering matrix that is employed in deriving the Faddeev equation: 
\begin{equation} 
[M_{qq}(k,q;K)]_{rs}^{tu} = \sum_{J^P=0^+,1^+,\ldots} \bar\Gamma^{J^P}\!(k;-K)\, \Delta^{J^P}\!(K) \, \Gamma^{J^P}\!(q;K)\,. \label{AnsatzMqq} 
\end{equation}  

One practical means of specifying the $\Gamma^{J^P}\!\!$ in Eq.\,(\ref{AnsatzMqq}) is to employ the solutions of a rainbow-ladder quark-quark Bethe-Salpeter equation (BSE).  Using the properties of the Gell-Mann matrices one finds easily that $\Gamma^{J^P}_C:= \Gamma^{J^P}C^\dagger$ satisfies exactly the same equation as the $J^{-P}$ colour-singlet meson {\it but} for a halving of the coupling strength \cite{regdq}.  This makes clear that the interaction in the ${\bar 3_c}$ $(qq)$ channel is strong and attractive.  For the correlations relevant herein, models typically give masses (in GeV) \cite{cjbsep,marisdq}:
\begin{equation}
\label{diquarkmass}
m_{(ud)_{0^+}} = 0.74 - 0.82 \,,\; m_{(uu)_{1^+}}=m_{(ud)_{1^+}}=m_{(dd)_{1^+}}=0.95 - 1.02\,.
\end{equation}
Such values are confirmed by results obtained in simulations of quenched lattice-QCD \cite{hess}.  

A solution of the BSE equation requires a simultaneous solution of the quark-DSE \cite{marisdq}.  However, since we have already chosen to simplify the calculations by parametrising $S(p)$, we also employ that expedient with $\Gamma^{J^P}\!$, using the following one-parameter forms: 
\begin{eqnarray} 
\label{Gamma0p} \Gamma^{0^+}(k;K) &=& \frac{1}{{\cal N}^{0^+}} \, 
H^a\,C i\gamma_5\, i\tau_2\, {\cal F}(k^2/\omega_{0^+}^2) \,, \\ 
\label{Gamma1p} {\tt t}^i \Gamma^{1^+}_\mu (k;K) &=& \frac{1}{{\cal N}^{1^+}}\, 
H^a\,i\gamma_\mu C\,{\tt t}^i\, {\cal F}(k^2/\omega_{1^+}^2)\,, 
\end{eqnarray} 
with the normalisation, ${\cal N}^{J^P}\!$, fixed by an appropriate analogue of Eq.~(\ref{canonicalnorm}) \cite{arneJ}. These {\it Ans\"atze} retain only that single Dirac-amplitude which would represent a point particle with the given quantum numbers in a local Lagrangian density: they are usually the dominant amplitudes in a solution of the rainbow-ladder BSE for the lowest mass $J^P$ diquarks \cite{cjbsep,marisdq} and mesons \cite{mr97,maristandyrho,pieterpion}. 

\subsubsection{Diquark propagators}
\label{qqprop}
Solving for the quark-quark scattering matrix using the rainbow-ladder truncation yields free particle propagators for $\Delta^{J^P}$ in 
Eq.~(\ref{AnsatzMqq}).  As already noted, however, higher order contributions 
remedy that defect, eliminating asymptotic diquark states from the spectrum.  The attendant modification of $\Delta^{J^P}$ can be modelled efficiently by simple functions that are free-particle-like at spacelike momenta but pole-free on the timelike axis \cite{mandarvertex}; namely,\footnote{These forms satisfy a sufficient condition for confinement because of the associated violation of reflection positivity.  This notion may be traced from Refs.\,\protect\cite{entire1,entire2,stingl,krein} and is reviewed in Refs.\,\protect\cite{cdragw,bastirev,alkoferrev}.}   
\begin{eqnarray} 
\Delta^{0^+}(K) & = & \frac{1}{m_{0^+}^2}\,{\cal F}(K^2/\omega_{0^+}^2)\,,\\ 
\Delta^{1^+}_{\mu\nu}(K) & = & 
\left(\delta_{\mu\nu} + \frac{K_\mu K_\nu}{m_{1^+}^2}\right) \, \frac{1}{m_{1^+}^2}\, {\cal F}(K^2/\omega_{1^+}^2) \,,
\end{eqnarray} 
where the two parameters $m_{J^P}$ are diquark pseudoparticle masses and 
$\omega_{J^P}$ are widths characterising $\Gamma^{J^P}\!$.  Herein we require additionally that
\begin{equation}
\label{DQPropConstr}
\left. \frac{d}{d K^2}\,\left(\frac{1}{m_{J^P}^2}\,{\cal F}(K^2/\omega_{J^P}^2)\right)^{-1} \right|_{K^2=0}\! = 1 \; \Rightarrow \; \omega_{J^P}^2 = \sfrac{1}{2}\,m_{J^P}^2\,,
\end{equation} 
which is a normalisation that accentuates the free-particle-like propagation characteristics of the diquarks {\it within} the hadron. 

\subsection{Nucleon and $\Delta$ Masses}
\label{NDmasses}
All elements of the Faddeev equations, Eqs.\,(\ref{FEone}) \& (\ref{FEDelta}), are now completely specified.  We solve the equations via the method described in Ref.\,\cite{oettelcomp}.  Owing to Eq.\,(\ref{DQPropConstr}), the masses of the scalar and axial-vector diquarks are the only variable parameters.  The axial-vector mass is chosen so as to obtain a desired mass for the $\Delta$, and the scalar mass is subsequently set by requiring a particular nucleon mass.  

Two primary parameter sets are presented in Table~\ref{ParaFix}.  Set~A is obtained by requiring a precise fit to the experimental nucleon and $\Delta$ masses.  It has long been known that this is possible; e.g., Ref.\,\cite{oettelfe} reports octet and decuplet baryon masses in which the rms deviation between the calculated mass and experiment is only $2$\%.  However, it is also known that such an outcome is undesirable because, e.g., studies using the cloudy bag model \cite{cbm} indicate that the nucleon's mass is reduced by as much as $\delta M_N = -300$ to $-400\,$MeV through pion self-energy corrections \cite{bruceCBM}.  Furthermore, a perturbative study, using the Faddeev equation, of the mass shift induced by pion exchange between the quark and diquark constituents of the nucleon obtains $\delta M_N = -150$ to $-300\,$MeV~\cite{ishii}.  This leads to Set~B, which was obtained by fitting to nucleon and $\Delta$ masses that are inflated so as to allow for the additional attractive contribution from the pion cloud \cite{hechtfe}. 

\begin{table}[t]
\begin{center}
\caption{\label{ParaFix} Mass-scale parameters (in GeV) for the scalar and axial-vector diquark correlations, fixed by fitting nucleon and $\Delta$ masses: Set~A provides a fit to the actual masses; whereas Set~B provides masses that are offset to allow for ``pion cloud'' contributions \protect\cite{hechtfe}.  We also list $\omega_{J^{P}}= \sfrac{1}{\surd 2}m_{J^{P}}$, which is the width-parameter in the $(qq)_{J^P}$ Bethe-Salpeter amplitude, Eqs.\,(\protect\ref{Gamma0p}) \& (\protect\ref{Gamma1p}):  its inverse is an indication of the diquark's matter radius.  Sets A$^\ast$ and B$^\ast$ illustrate effects of omitting the axial-vector diquark correlation: the $\Delta$ cannot be formed and $M_N$ is significantly increased.  It is thus plain that the axial-vector diquark provides significant attraction in the Faddeev equation's kernel.}
\begin{tabular*}{1.0\textwidth}{
l@{\extracolsep{0ptplus1fil}}c@{\extracolsep{0ptplus1fil}}c@{\extracolsep{0ptplus1fil}}
c@{\extracolsep{0ptplus1fil}} c@{\extracolsep{0ptplus1fil}}c@{\extracolsep{0ptplus1fil}}c@{\extracolsep{0ptplus1fil}}}
\hline
set & $M_N$ & $M_{\Delta}$~ & $m_{0^{+}}$ & $m_{1^{+}}$~ &
$\omega_{0^{+}} $ & $\omega_{1^{+}}$ \\
\hline
A & 0.94 & 1.23~ & 0.63 & 0.84~ & 0.44=1/(0.45\,{\rm fm}) & 0.59=1/(0.33\,{\rm fm}) \\
B & 1.18 & 1.33~ & 0.79 & 0.89~ & 0.56=1/(0.35\,{\rm fm}) & 0.63=1/(0.31\,{\rm fm}) \\\hline
A$^\ast$ & 1.15 &  & 0.63 &  & 0.44=1/(0.45\,{\rm fm}) &  \\
B$^\ast$ & 1.46 &  & 0.79 &  & 0.56=1/(0.35\,{\rm fm}) &  \\
\hline
\end{tabular*}
\end{center}
\end{table}

\section{Nucleon and $\Delta$ $\sigma$-terms}
\label{LResults}
\subsection{Analysis}
In order to calculate $\sigma_{N,\Delta}$ in the framework we have outlined it is necessary to know the variation with current-quark mass of the elements described in Sects.\,\ref{subsubsec:S}--\ref{qqprop}.  That is straightforward for the dressed-quark described in Sect.\,\ref{subsubsec:S} so long as we assume that the explicit $m$-dependence is dominant in the neighbourhood of the physical current-quark mass.  As we saw in connection with Eq.\,(\ref{sMEQ}), this assumption may underestimate the $m$-dependence.

The diquark propagators in Sect.\,\ref{qqprop} each involve a single mass-scale, and the response of the nucleon and $\Delta$ masses to changes in these mass-scales is much as one would expect if $m_{0^+}$ \& $m_{1^+}$ were true diquark masses.  This is apparent in Table \ref{ParaFix}.  For example, in changing from Set\,A to B, $M_N$ increases by $240\,$MeV when $(m_{0^+}+m_{1^+})$ is raised by $210\,$MeV; and the increase of $m_{1^+}$ by $50\,$MeV raises $M_\Delta$ by $100\,$MeV.  This is an appealing feature of the model.  

However, the connection should not be viewed as trivial because, in addition, these mass-scale parameters appear in the diquark Bethe-Salpeter amplitudes described in Sect.\,\ref{qqBSA}.  These amplitudes constitute the final element in the Faddeev equation kernel.  In principle their pointwise form will evolve with the current-quark mass and in our model, owing to Eq.\,(\ref{DQPropConstr}), that is effected through an explicit dependence on the mass-scale characterising the appropriate diquark's propagator.  This entails that the momentum-space width grows with increasing diquark mass-scale, a property which is consistent with the Bethe-Salpeter equation studies of Ref.\,\cite{mr97}.  It follows that the normalisation factor in Eqs.\,(\ref{Gamma0p}) and (\ref{Gamma1p}) becomes larger with increasing diquark mass.  That leads to a reduction in the quark-quark $\leftrightarrow$ diquark transition probability amplitude and thereby a decrease in the amount of attraction in the Faddeev equation's kernel.  This explains why, in changing from Set\,A to B, $M_N$ and $M_\Delta$ rise by more than just the increase in the diquark masses.

Diquark masses can unambiguously be defined and calculated in rainbow-ladder truncation, and that calculation would provide the variation of $m_{0^+}$ and $m_{1^+}$ with current-quark mass; viz., an estimate of the diquark $\sigma$-terms $\sigma_{0^+}$ and $\sigma_{1^+}$.  Although a realistic calculation of $m_{0^+}$ and $m_{1^+}$ is available \cite{marisdq}, the evolution of these masses with current-quark mass is not.  We have therefore studied this problem using the algebraic model of Ref.\,\cite{mn83}, which gives (in MeV)
\begin{equation}
\label{diquarkdm}
\sigma_{0^+} =21 \gtrsim \sigma_{1^+} = 20 \gtrsim \sigma_{\rho} = 17\,;
\end{equation}
i.e., an evolution with current-quark mass of a diquark's mass which is similar to that of the $\rho$-meson.  This being the case, we proceed by supposing that the current-quark mass dependence of $m_{0^+}$ and $m_{1^+}$ is well approximated by that of the $\rho$-meson; namely, we adapt the fit of Ref.\,\cite{marisvienna} as follows: 
\begin{eqnarray}
\label{scalarm}
m_{0^+}(m) & = & m^0_{0^+} + \sqrt{c_0 m} + c_1 m\,,\\
\label{vectorm}
m_{1^+}(m) & = & m^0_{1^+} + \sqrt{c_0 m} + c_1 m\,,
\end{eqnarray}
with $c_0 = 0.195\,$GeV, $c_1=1.90$, and (in GeV)
\begin{equation}
\begin{array}{c|c|c}
 & \mbox{Set~A} & \mbox{Set~B} \\\hline
m^0_{0^+} & 0.59 & 0.75 \\[0.3ex]
m^0_{1^+} & 0.80 & 0.85
\end{array}
\label{mqq}
\end{equation}
This assumption, too, will likely bias our calculation toward an underestimate of $\sigma_N$ and $\sigma_\Delta$.

Now that the response of the Faddeev equation's kernel to changes in the current-quark mass has been elucidated, the calculation of $\sigma_N$ and $\sigma_\Delta$ is straightforward.  It is only necessary to evaluate $M_N$ and $M_\Delta$ on a small domain of current-quark mass, centred on the physical value, Eq.\,(\ref{mcq}); viz., $4.5< m\,({\rm MeV}) < 5.5$, and interpolate.  The $\sigma$-terms follow from Eq.\,(\ref{sigmaFermion}) and our results are presented in Table~\ref{sigmaresults}.  It is notable that so long as Eq.\,(\ref{diquarkdm}) is valid, $\sigma_N \gtrsim \sigma_\Delta$.

\begin{table}[t]
\begin{center}
\caption{\label{sigmaresults} ``Quark core'' $\sigma$-terms for nucleon and $\Delta$ evaluated from  Faddeev equation solutions using Eq.\,(\protect\ref{sigmaFermion}).  All dimensioned results tabulated are reported in GeV.  NB.\ On the domain $4.5< m\,({\rm MeV}) < 5.5$, $M_N(m)$ and $M_\Delta(m)$ are linear.
The \emph{Set} label indicates which parameters from Table~\ref{ParaFix} were used in solving the Faddeev equation.  The row labelled Set\,B$_{\tilde\sigma}$ is described in connection with Eq.\,(\protect\ref{newdqsigma}).
Recall that absent an axial-vector diquark there is no $\Delta$.
%
For comparison, a value of $\sigma_{N}$ may be inferred from the isospin-even elastic $\pi N$ scattering amplitude and analyses yield: $\sigma_{N} \approx 45\,$MeV \cite{borasoy,sainio} or $\sigma_N \simeq 67 \pm 8\,$MeV \cite{workman,schweitzer}.  Moreover, analyses of data from simulations of two-flavour lattice-regularised QCD yield: $\sigma_{N} \approx 51\,$MeV \cite{wrightsigma}; $\sigma_N \sim 51\ldots\,54\,$MeV \cite{frink}; and $\sigma_{N} \approx 49\,$MeV and $\sigma_{\Delta} \approx 21\,$MeV \cite{ulfDelta}.  A chiral-quark model analysis gives $\sigma_N = 45 \pm 5\,$MeV and $\sigma_{\Delta}=32\pm 3\,$MeV \cite{lyubovitsky}.  Our best estimates are presented in Eq.\,(\ref{sigmaresults2}).
}
%
%
%
\begin{tabular*}{1.0\textwidth}{l@{\extracolsep{0ptplus1fil}}c@{\extracolsep{0ptplus1fil}}
c@{\extracolsep{0ptplus1fil}}
c@{\extracolsep{0ptplus1fil}}
c@{\extracolsep{0ptplus1fil}}}\\\hline
Set & $M_N$ & $M_{\Delta}$ & $\sigma_N$ & $\sigma_\Delta$ \\
\hline
A &  0.94 & 1.23 & 0.046 & 0.042\\
B &  1.18 & 1.33 & 0.047 & 0.042 \\\hline
A$^\ast$  & 1.15 & & 0.040 &  \\
B$^\ast$  & 1.46 & & 0.041 &  \\\hline
B$_{\tilde\sigma}$ &  1.18 & 1.33 & 0.057 & 0.051 \\\hline
\end{tabular*}
\end{center}
\end{table}

With the kernel we have employed, the Faddeev amplitudes describe a baryon's dressed-quark core.  Differences between the amplitudes obtained with Set~A and Set~B are exposed in calculations of the nucleons' electromagnetic form factors \cite{arneJ}.  It is apparent in Table~\ref{sigmaresults} that the values of the $\sigma$-terms are almost independent of whether Set~A or Set~B is used.  For the nucleon, the value is influenced much more by whether, or not, the axial-vector diquark is retained.  It is thus evident that $\sigma_N$ and $\sigma_\Delta$ are primarily determined by the evolution of the constituents' masses with current-quark mass, and that the pointwise response of the Faddeev amplitude affects the results very little.  %
In this connection we observe that the $\sigma$-term is the $Q^2=0$ value of a form factor and hence is determined by global rather than local properties of the distribution of bound state constituents.

Here, too, it is natural to ask whether meson-loop self-energy contributions materially affect the results in Table~\ref{sigmaresults}.  This may be addressed following Ref.\,\cite{hechtfe}, which explains the nature of the $\pi N$-loop corrections to the Faddeev equation nucleon mass.  A reliable estimate of this mass shift is
\begin{eqnarray} 
\delta M_N & = &-\,6\pi\, \frac{f^2_{NN\pi}}{m_\pi^2} \int\frac{d^3 
k}{(2\pi)^3} \, \frac{\vec{k}^2\,u^2(\vec{k}^2)}{\omega_\pi(\vec{k}^2) [ 
\omega_\pi(\vec{k}^2) + \omega_N(\vec{k}^2) - M_N]} \,,
 \label{deltaMppCBM} 
\end{eqnarray} 
where $f_{NN\pi}^2 = g_{NN\pi}^2 m_\pi^2/(16 \pi M_N^2)$, $\omega_\pi(\vec{k}^2) = \sqrt{\vec{k}^2+m_\pi^2}$, $\omega_N(\vec{k}^2) = \sqrt{\vec{k}^2+m_N^2}$, and, with a Pauli-Villars regularisation of the nucleon's pion-induced self-energy, 
\begin{equation} 
\label{uklimit} 
u(\vec{k}^2) = 1/(1+\vec{k}^2/\lambda^2)\,.
\end{equation} 
In Eq.\,(\ref{uklimit}), $\lambda$ is the pion-loop regularisation scale.  It is finite and nonzero in a realistic calculation because neither the pion nor the nucleon is pointlike.

To estimate the contribution to $\sigma_N$ from the mass shift in Eq.\,(\ref{deltaMppCBM}) we evaluated this expression using a fixed value of $(f_{NN\pi}/M_N)=(g_A/f_\pi)=13.6\,$GeV$^{-1}$, and values of $M_N$ and $m_\pi$ that vary with the current-quark mass, $m$; namely, we used the nucleon mass obtained from the Set\,B Faddeev equation at a given value of $m$, which is described by (in GeV)
\begin{equation}
M_N^B = 1.13 + 9.20\,m\,, \; m\in [0.0045,0.0055]
\end{equation}
and \cite{mrt98}
\begin{equation}
m_\pi^2 = -\,2\, m \, \frac{\langle \bar q q \rangle^\pi}{f_\pi^2}
\end{equation}
where, with the model of Sect.\,\ref{subsubsec:S}, the calculated values are \cite{valencedistn}: \mbox{$\langle \bar q q \rangle^\pi = (-0.25\,{\rm GeV})^3$} and $f_\pi=0.090\,$GeV.  This procedure gave (in GeV)
\begin{equation}
\label{sigmapiNres}
\begin{array}{l|ll}
\lambda & 0.30 & 0.40 \\\hline
\rule{0em}{3ex} \delta\sigma_N^{\pi N} & 0.0037& 0.0058  
\end{array}
\end{equation}
viz., a contribution to $\sigma_N$ from the $\pi N$ self-energy diagram of $\sim 4-6\,$MeV.  NB.\ We list results with values of the regularisation parameter, $\lambda$, that are typical of realistic analyses \cite{arneJ,ashley}.  The smaller value of $\lambda$ is favoured in our model \cite{arneJ}.  Moreover, the values in Eq.\,(\ref{sigmapiNres}) are unchanged if the Set\,A Faddeev equation results for $M_N$ are used.

Comparing rows 2 \& 4 and 3 \& 5 in Table~\ref{sigmaresults}, it is apparent that the $\pi N$-loop contribution to $\sigma_N$, Eq.\,(\ref{sigmapiNres}), possess the same sign and is similar in magnitude to that arising from the axial-vector diquark.  This is typical of the constructive interference between the pion cloud and axial-vector diquark correlations in the Faddeev equation.

The shift in the nucleon mass owing to a $\pi \Delta$ self-energy contribution has the same sign as that in Eq.\,(\ref{deltaMppCBM}) and is no larger in magnitude.   This suggests that the total correction to $\sigma_N$ from $\pi N$ and $\pi \Delta$ self-energy corrections is
\begin{equation}
\delta\sigma_N^{\pi N + \pi \Delta} \simeq 7\,{\rm MeV}.
\end{equation}
The shift in $M_\Delta$ owing to $\pi N$ and $\pi \Delta$ self-energy corrections is approximately one-half of the analogous correction to $M_N$ \cite{wrightANU}.  (Naturally, this fact lays behind the choice of Set\,B parameters in Table~\ref{ParaFix}.)  Hence we deduce 
\begin{equation}
\delta\sigma_\Delta^{\pi N + \pi \Delta} \simeq 3\,{\rm MeV}.
\end{equation}

The primary remaining source of uncertainty within our model is the assumption associated with Eqs.\,(\ref{scalarm}), (\ref{vectorm}), which entails $\sigma_{0^+} = \sigma_{1^+} = \sigma_\rho$.  To develop a notion of the error this may introduce we repeated the calculation of $\sigma_N$ and $\sigma_\Delta$ using the Set\,B Faddeev equation parameters with $\sigma_{0^+} = \sigma_{1^+} = 1.2\,\sigma_\rho$ and the  {\textit Ansatz}: 
\begin{equation}
\label{newdqsigma}
m_{qq}(m) = m_{qq}^0 + m\, \sigma_{qq}\,, 
\end{equation}
with $m_{qq}^0$ given in Eq.\,(\ref{mqq}).  This gave the results listed as Set\,B$_{\tilde\sigma}$ in Table \ref{sigmaresults}.

\subsection{Result}
Combining the elements discussed above we judge that, in the model employed, the best estimate of the $\sigma$-terms is 
\begin{equation}
\label{sigmaresults2}
\sigma_N \simeq 60\,{\rm MeV} \,,\; \sigma_\Delta \simeq 50\,{\rm MeV}\,.
\end{equation}

\section{Epilogue}
\label{LEpilogue}
We illustrated the connection between the $\sigma$-term and the Feynman-Hellmann theorem with an explicit calculation of $\sigma_\pi$ in the rainbow-ladder truncation of the Dyson-Schwinger equations.  Naturally, in the vicinity of the chiral limit we obtain the model-independent result: $\sigma_\pi = m_\pi/2$.  

While on the subject of mesons, we also calculated $\sigma_\rho$, for which the rainbow-ladder truncation likewise provides a reliable estimate.  \emph{A priori} that need not have been the case because \mbox{$\rho\to\pi\pi$} and $\rho\to\omega\pi$ self-energy terms noticeably reduce $m_\rho^2$.  However, while the contribution to $\sigma_\rho$ from each term is relatively large, they are of similar magnitude but opposite sign and therefore cancel to a large extent.  We find $\sigma_\rho\simeq m_\rho/34$, a value that is appreciably smaller than $\sigma_N$ \cite{ulfrho,allton}.  For the $\omega$-meson, on the other hand, there is little to cancel the contribution to $\sigma_\omega$ from $\omega\to\rho\pi$, which is three-times larger than that to $\sigma_\rho$ from $\rho\to\omega\pi$.  Hence $\sigma_\omega\simeq m_\omega/23$.  A comparison of these values with $\sigma_\pi$ highlights the importance of dynamical chiral symmetry breaking to light-hadron masses.

To calculate $\sigma_N$ and $\sigma_\Delta$ we employed a Poincar\'e covariant Faddeev equation, which describes baryons as composites of confined-quarks and -diquarks.  Two parameters appear in the model Faddeev equation.  They are the mass parameters of the scalar and axial-vector diquark correlations, and were fixed by fitting stipulated masses of the baryons.  The dependence of these diquark mass-scales on the current-quark mass was a key assumption in our calculation of the baryon $\sigma$-terms and the principal source of uncertainty.  

The Faddeev equation solution describes a baryon's ``quark core''.  This must be augmented in a consistent fashion by chiral-loop corrections.  Those chiral corrections increase $\sigma_N$ by $\lesssim 15$\% and $\sigma_\Delta$ by $\lesssim 10$\%.

Our analysis indicates that $\sigma_N$ is large compared with estimates based on chiral effective theory.  Nonetheless, our value of $\sigma_N \approx 0.06\,M_N$ is consistent with modern experimental evaluations \cite{workman,schweitzer}.  

It has been argued, based on a perturbative treatment of $SU(3)$-flavour symmetry breaking \cite{donoghue}, that a large value of $\sigma_N$ implies a large value for the ratio $y= 2 \langle p|\bar s s |p\rangle/\langle p|\bar u u +\bar d d|p\rangle$; where large means $y/2 \geq 0.2$.  While we have not studied this ratio, there are numerous examples of explicit calculations in which that inference is invalid because baryon masses exhibit a strong nonlinear dependence on the $s$-quark mass; e.g., Refs.\,\cite{lyubovitsky,jaffe,birse,Myhrer}.  

With this study we have gained insight into the current-quark mass dependence of hadron properties.  It is of use to gather our numerical results and reiterate them here in the form
\begin{equation}
\frac{\delta m_H}{m_H} = \frac{\sigma_H}{m_H} \frac{\delta \bar m}{\bar m} \,,
\end{equation}
where $\bar m$ is defined in Eq.\,(\ref{eq2}).  At the physical value of the light current-quark mass we have
\begin{equation}
\begin{array}{l|ccccc}
 & \pi & \rho & \omega & N & \Delta \\\hline
\rule{0em}{3.5ex}\displaystyle\frac{\sigma_H}{m_H} & 0.498 & 0.030 & 0.043 & 0.064 & 0.041\\
\end{array}
\end{equation}
From this point one may proceed directly to investigate the impact of our results on the analysis of experimental observations that relate to the variation of nature's fundamental ``constants.''

\begin{acknowledge}
We thank P.~Maris and P.\,C.~Tandy for constructive conversations.
This work was supported by: 
Department of Energy, Office of Nuclear Physics, contract no.\ W-31-109-ENG-38; 
\textit{Helmholtz-Gemeinschaft} Virtual Theory Institute VH-VI-041; 
the \textit{A.\,v.\ Humboldt-Stiftung} via a \textit{F.\,W.\ Bessel Forschungspreis}; 
and benefited from the facilities of ANL's Computing Resource Center.
\end{acknowledge}

\appendix
\section{Euclidean Conventions} 
\setcounter{section}{1}
\label{App:EM} 
In our Euclidean formulation: 
\begin{equation} 
p\cdot q=\sum_{i=1}^4 p_i q_i\,; 
\end{equation} 
\begin{equation}
\{\gamma_\mu,\gamma_\nu\}=2\,\delta_{\mu\nu}\,;\; 
\gamma_\mu^\dagger = \gamma_\mu\,;\; 
\sigma_{\mu\nu}= \sfrac{i}{2}[\gamma_\mu,\gamma_\nu]\,; \;
{\rm tr}_[\gamma_5\gamma_\mu\gamma_\nu\gamma_\rho\gamma_\sigma]= 
-4\,\epsilon_{\mu\nu\rho\sigma}\,, \epsilon_{1234}= 1\,.  
\end{equation}

A positive energy spinor satisfies 
\begin{equation} 
\bar u(P,s)\, (i \gamma\cdot P + M) = 0 = (i\gamma\cdot P + M)\, u(P,s)\,, 
\end{equation} 
where $s=\pm$ is the spin label.  It is normalised: 
\begin{equation} 
\bar u(P,s) \, u(P,s) = 2 M 
\end{equation} 
and may be expressed explicitly: 
\begin{equation} 
u(P,s) = \sqrt{M- i {\cal E}}\left( 
\begin{array}{l} 
\chi_s\\ 
\displaystyle \frac{\vec{\sigma}\cdot \vec{P}}{M - i {\cal E}} \chi_s 
\end{array} 
\right)\,, 
\end{equation} 
with ${\cal E} = i \sqrt{\vec{P}^2 + M^2}$, 
\begin{equation} 
\chi_+ = \left( \begin{array}{c} 1 \\ 0  \end{array}\right)\,,\; 
\chi_- = \left( \begin{array}{c} 0\\ 1  \end{array}\right)\,. 
\end{equation} 
For the free-particle spinor, $\bar u(P,s)= u(P,s)^\dagger \gamma_4$. 
 
The spinor can be used to construct a positive energy projection operator: 
\begin{equation} 
\label{Lplus} \Lambda_+(P):= \frac{1}{2 M}\,\sum_{s=\pm} \, u(P,s) \, \bar 
u(P,s) = \frac{1}{2M} \left( -i \gamma\cdot P + M\right). 
\end{equation} 
 
A negative energy spinor satisfies 
\begin{equation} 
\bar v(P,s)\,(i\gamma\cdot P - M) = 0 = (i\gamma\cdot P - M) \, v(P,s)\,, 
\end{equation} 
and possesses properties and satisfies constraints obtained via obvious analogy 
with $u(P,s)$. 
 
A charge-conjugated Bethe-Salpeter amplitude is obtained via 
\begin{equation} 
\label{chargec}
\bar\Gamma(k;P) = C^\dagger \, \Gamma(-k;P)^{\rm T}\,C\,, 
\end{equation} 
where ``T'' denotes a transposing of all matrix indices and 
$C=\gamma_2\gamma_4$ is the charge conjugation matrix, $C^\dagger=-C$. 
 
In describing the $\Delta$ resonance we employ a Rarita-Schwinger spinor to 
unambiguously represent a covariant spin-$3/2$ field.  The positive energy 
spinor is defined by the following equations: 
\begin{equation} 
\label{rarita}
(i \gamma\cdot P + M)\, u_\mu(P;r) = 0\,,\;
\gamma_\mu u_\mu(P;r) = 0\,,\;
P_\mu u_\mu(P;r) = 0\,, 
\end{equation} 
where $r=-3/2,-1/2,1/2,3/2$.  It is normalised: 
\begin{equation} 
\bar u_{\mu}(P;r^\prime) \, u_\mu(P;r) = 2 M\,, 
\end{equation} 
and satisfies a completeness relation 
\begin{equation} 
\frac{1}{2 M}\sum_{r=-3/2}^{3/2} u_\mu(P;r)\,\bar u_\nu(P;r) = 
\Lambda_+(P)\,R_{\mu\nu}\,, 
\end{equation} 
where 
\begin{equation} 
R_{\mu\nu} = \delta_{\mu\nu} I_{\rm D} -\frac{1}{3} \gamma_\mu \gamma_\nu + 
\frac{2}{3} \hat P_\mu \hat P_\nu I_{\rm D} - i\frac{1}{3} [ \hat P_\mu 
\gamma_\nu - \hat P_\nu \gamma_\mu]\,, 
\end{equation} 
with $\hat P^2 = -1$, which is very useful in simplifying the positive energy 
$\Delta$'s Faddeev equation. 


\end{document}